# Nonlinear switching dynamics in a photonic-crystal nanocavity

Yi Yu,[a] Evarist Palushani, Mikkel Heuck, Dragana Vukovic, Christophe Peucheret, Kresten Yvind, and Jesper Mørk

DTU Fotonik, Technical University of Denmark, DK-2800 Kongens Lyngby, Denmark.

**Abstract**

We measure the nonlinear switching dynamics of an InP photonic crystal nanocavity for different pulse energies and wavelengths relative to the cavity resonance and observe saturation of the switching contrast and broadening of the switching window. The effects are analyzed by comparison with nonlinear coupled mode theory and explained in terms of large dynamical variations of the cavity resonance. The results are important for applications in optical signal processing and provide insight into the nonlinear optical processes that govern the dynamics of the refractive index and the absorption.

**Introduction**

The resonant character of a cavity can be used for switching of an injected optical signal by dynamically changing the refractive index of the cavity. If the cavity has a high quality ($Q$) factor, the intensity of a (pump) field coupled to the cavity will be locally enhanced and may change the cavity resonance through nonlinear effects, allowing control of the transmission of another (signal) field, as demonstrated in Refs. 1-8. Ultra-compact integrated switches can be realized using photonic-crystal (PhC) membranes, where a line-defect waveguide, or even a pair of crossing waveguides[9], is coupled to a point-defect nanocavity. Exploiting the small-modal volume and high $Q$-factor achievable in so-called H0-type cavities[10], the switching energy may be lowered to the femto-joule level[2]. Recent studies[4,11] identified the physical processes governing the switching dynamics, with free carrier effects playing a dominant role. In this work we show that the dynamics of the structure qualitatively changes with pump energy, with consequences for its application and optimization as an all-optical switch. The measured dynamics are shown to be accurately modeled by nonlinear coupled mode theory and the dynamical effects giving rise to the complex switching dynamics are explained.

---

[a] Electronic mail: yiyu@fotonik.dtu.dk

**Pump-probe experiments**

As shown in the scanning electron microscope (SEM) image in Fig. 1(a), the investigated structure is an air-slab PhC membrane based on InP with a lattice constant, hole radius and membrane thickness of 425 nm, 98 nm and 340 nm, respectively. It consists of an H0-type cavity formed by shifting two neighboring air holes 85 nm in opposite directions, giving a resonance with an intrinsic $Q$-factor of $\sim 2.5 \times 10^4$ and mode volume of $0.22(\lambda/n)^3$, as estimated through finite-difference time-domain calculations. The cavity is coupled to in- and output waveguides (WGs) of the standard W1-defect type, albeit with the two innermost arrays of holes being shifted towards the WG center by 42.5 nm to make the waveguide fundamental mode overlap in frequency with the cavity mode. The device is fabricated using a combination of electron-beam lithography, reactive-ion etching and selective wet-etching. For detailed information about the fabrication process, see Ref. [9].

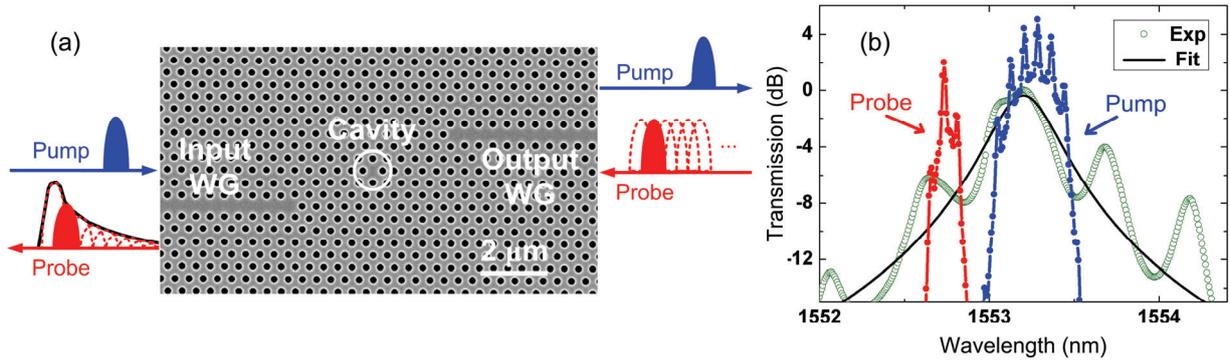

Fig. 1. (a) SEM image of fabricated InP membrane structure and illustration of pump-probe measurement technique. (b) Input pump (blue dotted line) and probe (red dotted line) spectra relative to the unperturbed cavity resonance spectrum (green circles), which is fitted with a Lorentzian line (black solid line) corresponding to a loaded $Q$-factor of $4\times10^3$.

A homodyne pump-probe technique is employed to investigate the switching dynamics[4]. A train of ~1.5 ps (FWHM) pulses, at a repetition rate of 1.25 GHz, are carved out from a 10 GHz pulse train using an intensity modulator. Pump and probe signals (both TE-polarized) with an adjustable temporal delay are generated by band-pass filtering using a multi-port processor (WaveShaper 4000S). A counter-propagating scheme, cf. Fig. 1(a), is employed to avoid pump-probe interactions in the waveguide sections of the device. The transmitted probe signal is selected using a band-pass filter and then detected with a photodiode, amplified, and monitored with a lock-in amplifier. When the pump spectrum overlaps with the cavity resonance, free carriers are generated through two-photon absorption, which blue shift the cavity resonance[2-4] via plasma effects and band-filling. By measuring the output probe energy as a function of time delay, cf. illustration in Fig. 1(a), the switching dynamics can be monitored. For a probe that is initially blue-detuned with respect to the cavity (switch-on case), the transmission of the probe therefore first increases and subsequently decreases, reflecting a temporal regime of fast carrier diffusion followed by slow carrier recombination[4,11].

Fig. 1(b) shows the transfer function of the cavity, displaying a resonant wavelength at 1553.2 nm with a loaded $Q$-factor of $4\times10^3$ corresponding to a photon lifetime of 3 ps. The pump has a spectral width (FWHM) of 0.3 nm and is 0.05 nm red-detuned from the resonance (i.e. towards longer wavelengths), while the probe, with a spectral width of 0.1 nm, is tuned to the blue side of the resonance. The discrete lines seen in the pulse spectra, with a periodicity of about 0.08 nm, originate from the 10 GHz light source from which the 1.25 GHz pulse trains are obtained by suppression of 7 out of 8 pulses using an external intensity modulator with finite extinction ratio. We found that these small ripples can to some extent decrease the switching contrast and elongate the switching time, but not significantly. The ripples observed in the cavity spectrum originate from reflections at the WG facets and the interfaces of cavity in the device.

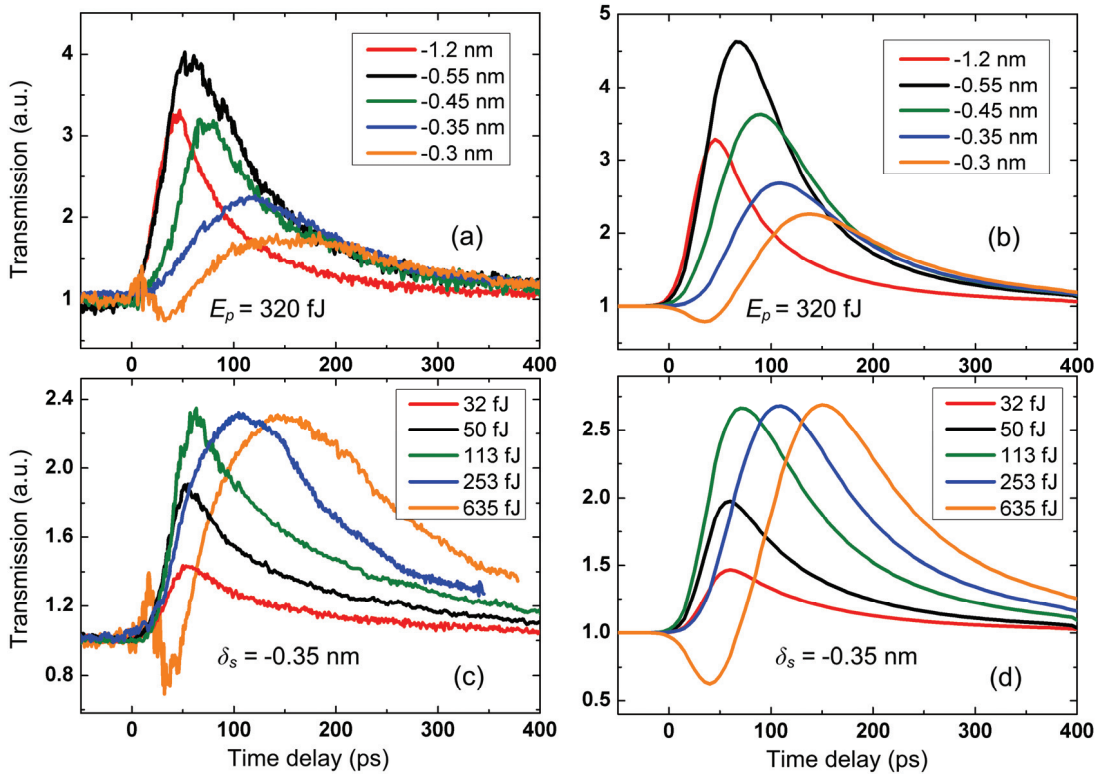

Fig. 2. (a) Measured and (b) simulated probe transmission versus pump-probe delay for fixed pump energy $E_p$ of 320 fJ and different values of probe detuning $\delta_S$. (c) Measured and (d) simulated probe transmission for fixed probe detuning $\delta_S$ of -0.35 nm and different pump energies. The transmission is normalized to one when the probe precedes the pump.

Fig. 2(a) shows measured pump-probe traces with different probe detunings for a fixed pump energy of 320 fJ, estimated by subtracting the coupling loss (~7.5 dB) from the measured pump power at the input fiber. For a detuning of -1.2 nm, the expected and previously observed dynamics[11] is observed. As the probe moves closer to the cavity resonance, the switching contrast increases, since the Lorentzian transmission spectrum has a larger slope close to the resonant peak than in the tail. However, when the probe moves even closer to the cavity resonance (from -0.55 nm to -0.3 nm), the switching contrast decreases and at the same time, the switching time-window broadens. In addition, one observes an initial decrease of the transmission right after the pump pulse excitation. Qualitatively similar behavior is observed in Fig. 2(c), where the probe detuning is kept fixed at -0.35 nm, while the switching energy is varied. Upon increasing the switching energy, the switching contrast first increases and then saturates accompanied by a significant broadening of the switching window. For the largest pulse energy, one also observes an initial transmission decrease, below the level prior to pump excitation.

**Theory**

In order to understand the physical mechanisms responsible for the dynamics observed experimentally, we perform simulations based on a temporal coupled mode theory[4,11-14]:

$$\dot{a}_x(t) = \left( -(\gamma_v + \gamma_{in} + \gamma_{TPA}^x(t) + \gamma_{FCA}^x(t))/2 - j(\omega_o + \Delta\omega_{c,x}(t) - \omega_x) \right) a_x(t) + \sqrt{\frac{\gamma_{in}}{2}} s_x^i(t), \quad x = p, s \quad (1)$$

Here, $a_x(t)$ is the complex cavity field, with $|a_x(t)|^2$ representing the energy in the cavity within the bandwidth of the pump ($x=p$) and signal ($x=s$) pulse, respectively, and $|s_x^i(t)|^2$ denote the injected probe and pump power. The cavity has an unperturbed resonant frequency of $\omega_o$ and the center pump and probe frequencies are denoted $\omega_x$. The loss rate $\gamma_{in}$ represents cavity-waveguide coupling, while $\gamma_v$ represents coupling to all other modes but the relevant waveguide modes, mainly vertical emission. The refractive index at the local position of the cavity is changed by the pump field, considered much stronger than the signal, due to the Kerr effect, the dispersion of free carriers generated by the field, as well as thermal effects. The corresponding modulation of the cavity resonance is modelled as[4,11]

$$\Delta\omega_{c,x}(t) = -K_{Kerr}^{x}|a_p(t)|^2 + K_{Car}N(t) - K_{th}\Delta T(t),$$ with $K_{Kerr}^{x}$, $K_{Car}$ and $K_{th}$ being constants. $K_{Kerr}^{s}$ differs from $K_{Kerr}^{p}$ by a factor of 2 due to the different magnitudes of cross- and self-phase modulation. The mode-averaged carrier density $N(t)$ and temperature change $\Delta T(t)$ are governed by rate equations that can be found in Ref. 4. This model takes into account the joint effects of fast carrier diffusion and slow surface and bulk recombination via three time constants and has shown excellent agreement with experiments[4]. In the case considered here, where an InP membrane is excited at wavelengths around 1.5 μm, carrier excitation occurs only by two-photon absorption and this effect, as well as free-carrier absorption, also changes the loss rates, represented by the terms $\gamma_{TPA}^{x}(t)$ and $\gamma_{FCA}^{x}(t)$ [4,11]. The output probe field is

$$s_s^o(t) = \sqrt{\frac{\gamma_{in}}{2}}a_s(t) \qquad (2)$$

The transmission dynamics measured using the lock-in amplifier is proportional to the output probe pulse energy $E_s(\tau_d) = \int_t |s_s^o(t,\tau_d)|^2 dt$ and depends on the time delay $\tau_d$ between the probe and pump.

Figs. 2(b) and 1(d) show the simulated results using parameter values corresponding to the experimentally investigated structures. Other material parameters may be found in Ref. 4. The fast and slow carrier relaxation times are extracted to be approximately 4 ps, 85 ps and 250 ps, respectively. The simulated results show good qualitative agreement with the measurements and, in particular, account for the switching window broadening and temporal dip.

In order to further elucidate the dynamics, we consider the situation where the probe is a weak continuous wave (CW) beam, thus avoiding the blurring of the cavity resonance dynamics associated with a probe pulse that is short or comparable to the cavity decay time. Under CW probing, we perform the same set of simulations as shown in Fig. 2(d) for a pulsed probe signal, i.e. fixing the probe detuning but varying the pump energy. We found that the physical mechanism responsible for the saturation broadening effect shown in Figs. 2(a) and 2(c) are the same.

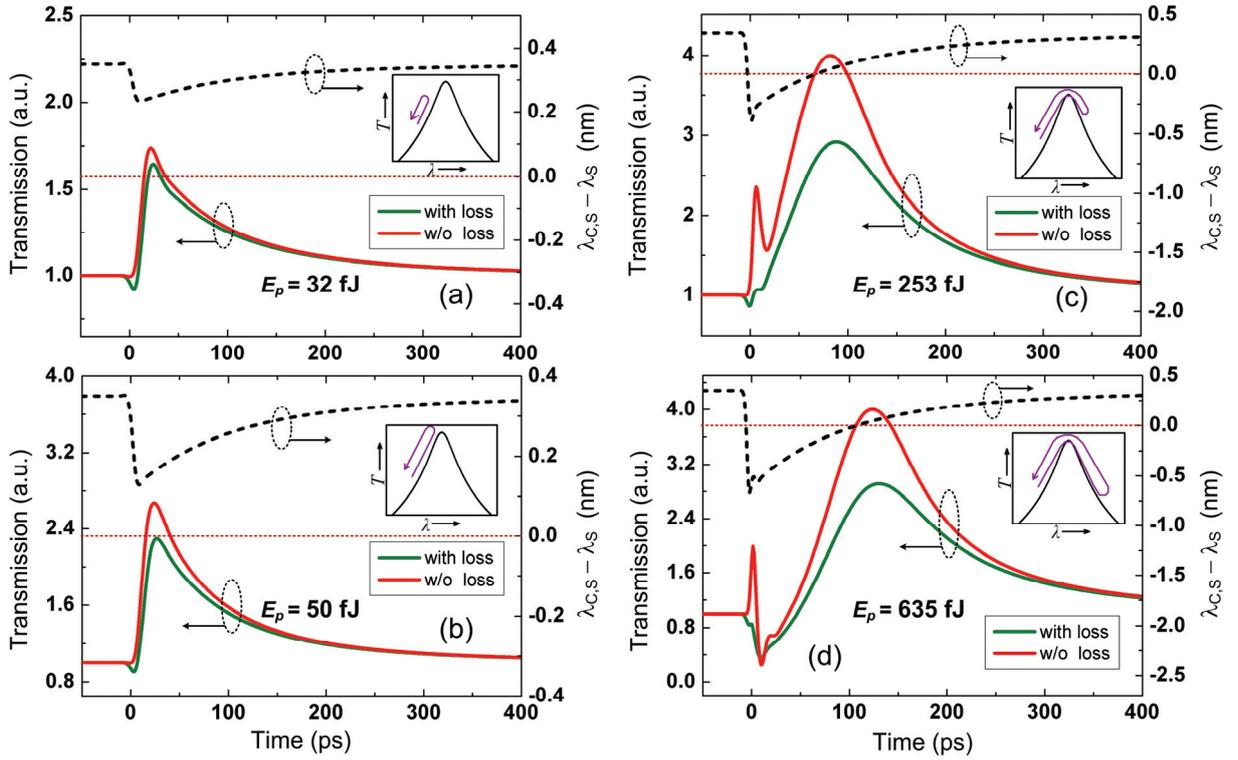

Fig. 3. Simulated output probe power (left axis, normalized to 1 for large negative time delay) versus time and perturbed resonance wavelength (right axis) for fixed initial detuning of -0.35 nm and pump energies $E_p$ of (a) 32 fJ, (b) 50 fJ, (c) 253 fJ and (d) 635 fJ. At the intersection(s) of the horizontal red dotted line and the black dashed line, the cavity resonance shifts across the probe wavelength, i.e. $\lambda_{c,s} - \lambda_s = 0$. The red (green) trace corresponds to the case where the pump-induced losses are excluded (included). The insets illustrate the temporal evolution of the probe transmission relative to the cavity transmission spectrum.

Fig. 3 shows the corresponding simulated variation of the output probe power $\left|s_s^o(t)\right|^2$ for an input probe wavelength, $\lambda_s$, that is blue detuned by 0.35 nm from the unperturbed cavity resonance. The other parameters are the same as in Fig. 2(d). The dynamical detuning between the perturbed cavity resonance wavelength $\lambda_{c,s}$ and $\lambda_s$ is also plotted (black dashed lines). Two different simulations are carried out: green solid lines are for the case where the pump-induced losses are included ($\gamma_{\text{TPA}}^s(t) \neq 0$, $\gamma_{\text{FCA}}^s(t) \neq 0$), while the red lines are for the case, where dynamical losses are neglected ($\gamma_{\text{TPA}}^s(t) = 0$, $\gamma_{\text{FCA}}^s(t) = 0$). In the latter case, the probe dynamics is only governed by the resonance shift.

As seen by comparing Figs. 3 and 2, the temporal variation imprinted on the CW input probe beam by the pump pulse are similar to results observed when both pump and probe beams are pulsed and the delay between them is varied. For relatively low pump energies, Figs. 3(a) and 3(b), the pump-induced losses are seen to play only a minor role. For increasing pump energy, Figs. 3(c) and 3(d), the maximum transmission contrast saturates, accompanied by the appearance of two local maxima for the case where nonlinear losses are neglected (red solid lines). From the corresponding temporal variation of the detuning, $\lambda_{c,s} - \lambda_s$, these maxima are seen to appear at the temporal positions where the dynamically changing cavity resonance matches the probe wavelength, $\lambda_{c,s} - \lambda_s = 0$. If the cavity resonance shifts beyond the probe wavelength, the probe and cavity wavelengths match twice during every switching event, and the latter maximum appears during the decay and is thus delayed by an amount depending on the refractive index change induced by the pump. When nonlinear losses are accounted for, green lines in Figs. 3(c) and 3(d), the first maximum is seen to be suppressed mainly due to two-photon absorption, finally forming an output pulse shape that is delayed and broadened in time, in agreement with the experimental observations. In addition, the pump energy of 635 fJ, Fig. 3(d), induces a red detuning which is larger in absolute magnitude than the original blue detuning, cf. the inset of Fig. 3(d), thus reducing the probe transmission below its initial state, in agreement with the experimental results in Fig. 2(c).

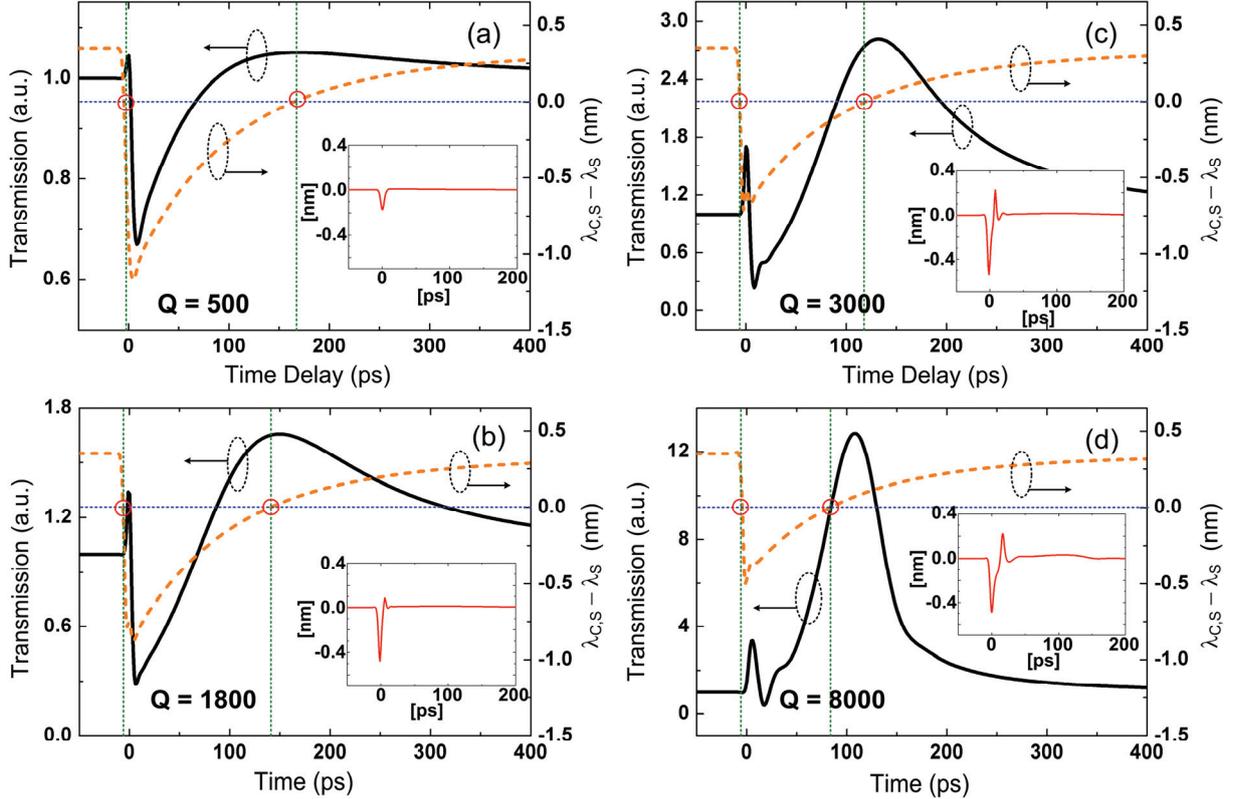

Fig. 4. Simulated probe transmission (black solid lines) and corresponding dynamical detuning $\lambda_{c,s} - \lambda_s$ (yellow dashed lines) for a CW probe signal. The pump energy is 635 fJ, the initial detuning is -0.35 nm and the cavity has a loaded $Q$-factor of (a) 500, (b) 1800, (c) 3000 and (d) 8000. The red circles indicate the time at which the resonance passes across the probe. Nonlinear losses are excluded for the probe. The insets show the temporal variations of the offset of the output probe wavelength with respect to the input probe wavelength $\lambda_s$.

We notice that, although the two transmission maxima displayed by the red solid lines in Figs. 3(c) and 2(d) both correspond to the probe coinciding with the peak of the cavity transmission, the corresponding levels of transmission largely differ, even when nonlinear losses are neglected. To further characterize this behavior, we calculated the probe transmission dynamics for structures with different (loaded) $Q$-factors and all other parameters kept fixed, cf. Fig. 4.

For the structure with a low $Q$-factor of 500, Fig. 4(a), the two maxima correspond to almost the same transmission level. However, as the $Q$-factor increases the transmission observed at the second maximum increases beyond the first one (Figs. 4(b)-4(d)). This can be explained from a cavity energy point of view since the output probe power is directly linked to the cavity energy through Eq. (2). For low $Q$-factors, Fig. 4(a), where the cavity life time is short compared to the pump pulse width (~12 ps), the cavity energy, and thus the transmitted probe intensity, cf. Eq. (2), follows the dynamical detuning, and the transmission levels at the two maxima are approximately the same. For larger $Q$-factors, however, the "charging" time of the cavity is longer than the pulse width, and the first local transmission maximum does not reach the value that would be expected under stationary conditions. On the other hand, the relaxation of the resonance back to its initial value happens at a much slower rate, governed by the carrier relaxation, and the probe transmission maximum reaches the value expected under stationary conditions. Besides the difference in transmission levels, the time delay between the maxima also increases with $Q$-factor, which is also well understood since the higher the $Q$-factor is, the slower the response of the cavity energy becomes. In addition, from Fig. 4, it is interesting to see that the nonlinear resonance shift is larger for a cavity with low $Q$-factor compared to a cavity with high $Q$-factor. This reflects that for a cavity with high $Q$-factor, a smaller amount of pump energy would be coupled into the cavity. Besides, the resonance may shift considerable compared to the spectrum of the pump pulse, thus lowering the efficiency with which pump energy is converted into refractive index change.

Insets in Fig. 4 illustrate how the wavelength of the transmitted probe signal also changes with time due to the temporal variation of the refractive index induced by the pump signal. The frequency shift of the probe signal can be obtained using $\Delta\Omega_s(t) = -\partial\phi_s(t)/\partial t$ where $\phi_s(t)$ is the phase of the slowly varying envelope of the cavity mode, $a_s(t)$. We observe that $\Delta\Omega_s(t)$ differs significantly from the cavity resonance shift, $\Delta\omega_{c,s}(t)$, represented by the yellow dashed curve ($\lambda_{c,s} - \lambda_s$). In the case where the cavity resonance is perturbed on a time scale shorter that the cavity lifetime and the cavity field subsequently relaxes *without external injection*, i.e. the feeding term in Eq. (1) is absent, the frequency of the output signal follows the cavity shift, $\Delta\Omega_s(t) = \Delta\omega_{c,s}(t)$. This chirping process, termed adiabatic wavelength conversion[15], has been observed for cavities with relatively large $Q$-factors[16-18]. In the case considered here, where there is a continuously injected probe signal and the perturbation of the cavity resonance happens on a time scale shorter compared to the cavity photon lifetime, we find that the process rather should be described as cross-phase modulation and the probe frequency variation, $\Delta\Omega_s(t)$, caused by the pump pulse, rather follows the time derivative of $\Delta\omega_{c,s}(t)$, as seen in Fig. 4. For very small pump-probe detuning (-0.3 nm in Fig. 2(a)) or very large pump energy (635 fJ in Fig. 2(c)), we find that the dynamical wavelength shift of the pump is large enough to drag part of it into the pass-band of the filter centered around the probe wavelength, causing interference effects that are observed in the experiments as an oscillatory pattern around zero time delay in the detected probe traces, cf. Figs. 2(a) and 2(c).

**Conclusion**

In summary, we studied the nonlinear switching dynamics of InP photonic-crystal nanocavities and found that the response undergoes qualitative changes when the pump energy, detuning or cavity quality factor are varied. Transmission saturation, broadening of the switching window as well as an initial temporal reduction of the transmission were observed. Good agreement was found between experimental results and simulations using coupled mode theory, based on which the observations could be explained as a consequence of the dynamical evolution of the cavity resonance frequency in combination with nonlinear losses and the time needed to establish the cavity field.

**Acknowledgements**

The authors acknowledge financial support from Villum Fonden via the NATEC (NAnophotonics for TErabit Communications) Centre.